# Novel BCI paradigm for ALS patients based on EEG and Pupillary Accommodative Response

DAVIDE D'ADAMO, EMILIANO ROBERT, and CRISTINA GENA, Department of Computer Science, University of Turin, Italy

SILVESTRO ROATTA, Neuroscience Department "Rita Levi Montalcini", University of Turin, Italy

ABSTRACT. Brain-computer interfaces (BCIs) are one of the few alternatives to enable locked-in syndrome (LIS) patients to communicate with the external world, while they are the only solution for complete locked-in syndrome (CLIS) patients, who lost the ability to control eye movements. However, successful usage of endogenous electroencephalogram(EEG)-based BCI applications is often not trivial, due to EEG variations between and within sessions and long user training required. In this work we suggest an approach to deal with this two main limitations of EEG-BCIs by inserting a progressive and expandable neurofeedback training program, able to continuously tailor the classifier to the specific user, into a multimodal BCI paradigm. We propose indeed the integration of EEG with a non-brain signal: the pupillary accommodative response (PAR). The PAR is a change in pupil size associated with gaze shifts from far to close targets; it is not governed by the somatic nervous system and is thus potentially preserved after the evolution from LIS to CLIS, which often occurs in neurodegenerative diseases, such as amyotrophic lateral sclerosis. Multimodal BCIs have been broadly investigated in literature, due to their ability to yield better overall control performances, but this would be the first attempt combining EEG and PAR. In the context of the BciPar4Sla, we are exploiting these two signals, with the aim of developing a more reliable BCI, adaptive to the extent of evolving together with the user's ability to elicit the brain phenomena needed for optimal control, and providing support even in the transition from LIS to CLIS.

## 1 INTRODUCTION

Locked-In Syndrome (LIS) is a rare neurological condition, possibly due to neurodegenerative diseases, such as amyotrophic lateral sclerosis (ALS), which bring the patient to the incapability of any voluntary movement, except for eye movements. A LIS patient is therefore unable to communicate autonomously, and a set of solutions exists to assist communication in this population: from no-tech (e.g. E-Tran boards) to high-tech (e.g. eye-tracker based systems), these solutions are mainly based on residual muscular control or eye-gaze [32].

One way to assess LIS patients' intentions without depending on eye movements is relying on brain signals. The most common approach to non-invasively extract information about brain activity is through EEG signals, which represent the activity of neuronal ensembles in cortical areas underneath the electrodes [23]. A Brain-Computer Interface (BCI) [22, 31] is a system capable of using this kind of signals to control any kind of external effector, from robotic arms to communication procedures. The main brain phenomena exploited in BCI control are visually evoked potentials [3, 9], when used in speller systems and necessarily phase-locked to an external stimulus (exogenous BCI), and sensory-motor rhythms (SMR) variations, spontaneously evocable by the user (endogenous BCI) by practicing Motor Imagery tasks (MI), i.e. imagined movement of a specific body-part [36]. Sadly, EEG-based BCIs suffer from different issues: low signal to-noise ratio, high inter-session instability, need of periodical recalibration of the predicting model and, in case of SMR control, long user-training process [17, 21, 23, 29, 34].

From this point of view, a BCI could benefit from a secondary input other than EEG, reliable and easy to control by the user, to support him/her during the MI training period giving the possibility to control the interface since day zero. Pupillary accommodative response (PAR) [8, 25], is a good candidate for this role: the control signal is elicited by the natural act of gaze shifting from a far to a near target, making the learning process for the user quite straightforward. PAR is based on the variations of pupil size, governed by the autonomic nervous system; therefore it is potentially retained by complete locked-in syndrome (CLIS) patients, which represent the subsequent stage in many cases of LIS (e.g. when caused by neurodegenerative diseases), defined as when the patient also loses the control of the eye movements. To our knowledge, this is the first attempt in literature to develop a multimodal endogenous BCI combining EEG and PAR.

The idea presented in this work aims at developing a low-cost adaptive EEG-based BCI, easy to use at the beginning thanks to PAR support, capable of continuously improving both the BCI classification model and user's ability in controlling SMR and providing a safer transition to CLIS conditions, usually an obligated passage for this kind of patients. The work has been realized in context of the BciPar4Sla, which is a follow-up of past project on BCI [13] and aims to develop an innovative form of human-machine interaction based on two possible communication channels: brain waves voluntarily modulated by the patient (EEG) and pupillary movement.

This paper has been organized as follows: Section 2 discusses related work in the field, Section 3 presents our approach, while Section 4 concludes the paper.

## 2   RELATED WORK

PAR signal has been proven robust and effective in terms of human-computer interface control [25] and allowed to develop a stand-alone Augmentative and Alternative Communication device, e-Pupil [6], enabling the user to answer simple questions or summon the attention of caregivers, yielding an accuracy of 100ti over a 4 class discrimination paradigm, based on duration and instant of initiation of pupillary constriction.

Integrating an EEG-based BCI with such a control signal would make it a multimodal BCI [20], potentially yielding better performance in target detection and/or allowing multidimensional control. Many example of multimodal BCI can be found in literature. For example, Kim et al. combined mental states recognition and eye-gaze direction to increase the range of commands callable by the user in a quadcopter driving task, while keeping the UI intuitive and simple. Another example would be the work of Pfurtscheller et al., in which it is demonstrated that using a MI-based switch to activate a steady state visual evoked potential BCI paradigm helps substantially to reduce the misclassification rate. In this way an exogenous paradigm could be used in a self-paced manner, exploiting the endogenous nature of MI. Finally, de'Sperati et al. combined pupillary frequency tagging, due to pupillary response to light intensity periodical oscillations, and steady state visual evoked potential to increase accuracy in a simple binary communication protocol, but to our knowledge there is no work in literature using both PAR and MI in a multimodal BCI paradigm.

In order to exploit SMR as control signals the user needs to learn how to correctly execute MI tasks, and to this aim usually neurofeedback (Nti) training protocols are adopted [10, 19, 27]. Nti training sessions consist in short trials (less than 20 seconds), in which the user is told what MI task to exercise (e.g. right hand, left hand, feet) and given a feedback somehow linked to the online classification score, enabling him/her to hone the execution technique and reach better performances (i.e. faster SMR modulation, better separation between classes). Being the user progressively learning how to elicit changes in SMR, training the classifier only at the beginning of the BCI experience (i.e. before the user finds the best way to execute MI) is not an efficient strategy. Therefore Nti training embedded with online model adaptation techniques emerged as effective tools for BCI users in order to obtain optimal BCI performances [1, 11, 14].

## 3   APPROACH

This section describes the development perspectives of the current work: subsection 3.1 introduces the signals exploited as controls over the interface, their generation, the required processing and the classification methods; subsection 3.2 describes the User Interface (UI) in terms of interaction modalities and adaptation to user's control capabilities; finally, subsection 3.3 describes the user training protocol.

## 3.1 Control Signals

The BCI to be developed will be based on two main physiological signals: pupil area and SMR variations, obtained respectively through PAR and MI tasks. In the following paragraphs signal generation, acquisition, processing and classification for both phenomena are briefly described.

*PAR: Task Execution.* Pupil constriction is obtained due to PAR when the user shift the gaze from a far target to a near one. A single PAR task is executed shifting the gaze to the near target, and back to the far one. Taking inspiration from previous works exploiting this phenomenon [5, 6, 25, 33], the main UI display will constitute the far target (about 150 cm from the subject), while a transparent plastic sheet covered in white dots placed about 30 cm away from the subject will constitute the near target. The acquisition device will be an infrared (IR) camera, coupled with an IR LED, mounted on a customized pair of eyeglasses and connected to the PC, although the intended application could in principle work well with a remote eye-tracking system. However, most remote eyetrackers do not provide a real-time access to pupil size measurement and may be considerably expensive. On the contrary, the present prototype was developed following a low-cost approach, which however grants full control of all acquisition and processing steps [5, 6, 25], as required in research applications.

*PAR: Image Processing and Classification.* The image processing pipeline to be adopted reproduces the one presented in [5]. Briefly, after a preliminary automatic identification of the region of interest (ROI) containing the pupil, frames coming from the camera are cropped to match the ROI, processed via the ellipse from method described in [28] to detect the pupil, whose area can finally be computed. Signal conditioning applied to the pupil area time series follows the pipeline designed in [6], and allows to cope with physiological fluctuation of pupil size.

*MI: Task execution.* As said above, SMR modulation can happen as a consequence of specific mental tasks: during MI, the user performs an imaginary movement of a specific body part, which will trigger, similarly to an actual motor action, a frequency- and location-specific modulation of the EEG power (event related (de)synchronization; for details see [16]). The EEG acquisition setup consists of an electrodes headset, a bioamplifier and a processing workstation (PC). Two different acquisition system will be tried in this work: OpenBCI[1] CytonDaisy board (bioamplifier) coupled with a

Greentek[2] EEG cap (headset), and an Emotiv[3] EPOC+ (embedded). Proceeding with implementation and testing the best performing system will be chosen.

*MI: EEG Classification of mental tasks.* Among the different methods designed in the last decades, according to literature [21, 35] Riemannian Geometry based classifiers are considered state-of-the-art for MI tasks classification. following the pipeline described in [11], for the initial model training, the EEG data is bandpass filtered and divided in 50ti overlapping 0.5 s long labeled epochs, covariance matrices are computed and averaged in the Riemannian space for the different classes (at least one MI task and the idle state, i.e. no-control) obtaining class-specific prototypes. During actual classification, new data epochs will be classified based on the Riemannian distance from class prototypes.

---

[1] http://www.openbci.com/
[2] https://www.greenteksensor.com
[3] https://www.emo8v.com/

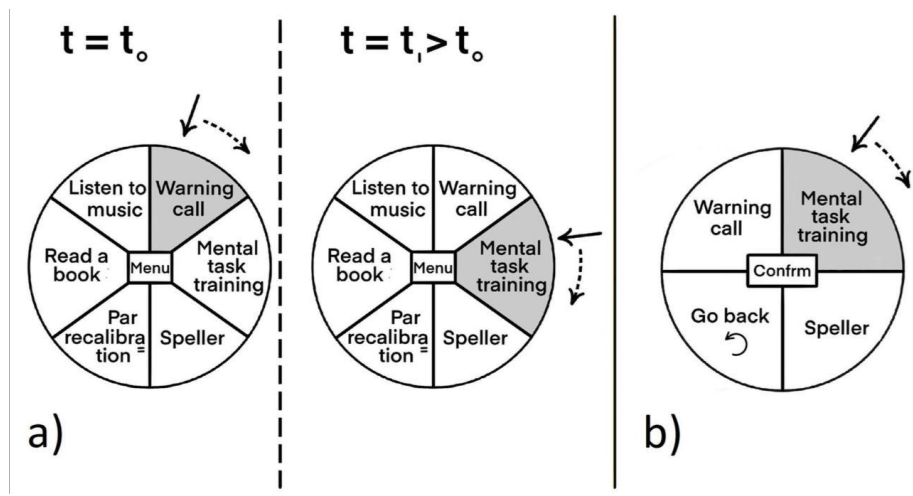

fig. 1. a) main menu view; the solid arrow glides along the edge of the menu highligh8ng one by one the choices; selec8on is achieved by execu8ng a PAR task. b) confirma8on view; the only choices available are the one ini8ally selected, the two surrounding the laSer and the *go-back* op8on.

*Model adaptation.* To address EEG strong non-stationarity [23] adaptive approaches are encouraged [29], indeed their classifier includes a comparison of strategies to continuously update the classifier references (class prototypes in the Riemannian space). The best performing in terms of classification accuracies and computational cost is based on a periodical re-estimation of the Riemannian class means considering both incoming new data epochs and the previous prototypes, the later being heavily weighted. Moreover, the classifier adaptation takes place during Nti training sessions, matching perfectly the intents of our work. Therefore the adaptation method to be implemented will strongly take inspiration from the just presented design.

## 3.2 User interface

Another core objective of this work is developing a BCI application able of adapting to the user in a smart and personalized way [4], helping him/her to make his/her *preferential choices* [15]. This aim can be reached designing an adaptive user interface (UI), able to evolve together with the control capabilities of the user. At the beginning, when the user is not confident with the execution of MI yet, the interface can be driven using PAR only (PAR-based UI). After an initial period of user MI training through Nti protocols, and automatic model fine-tuning, if the scores obtained in the Nti training sessions are high enough (see subsection 3.3), the UI evolves to a stage where both PAR and EEG can be exploited to obtain a smoother user experience. In any of these configurations, the possibility of going back to the previous menu must always be present and the UI should give the possibility to i) easily access Nti training sessions and ii) promptly call the caregiver when needed. In the following paragraphs the design of interaction for the PAR-based and multimodal configurations is described, and finally the possibility of answering simple external questions is discussed.

*PAR-based UI.* When the UI has to be entirely driven by PAR, the main paradigm for taking choices is selecting them from a dynamic menu by executing the task, and confirming the selection in a secondary *confirmation menu*. In this way, executing two PAR tasks the user achieves a successful selection, supposedly in less than five seconds [6]. A plausible example of main menu is presented together with an example of confirmation menu for the choice "Mental task training" (Fig. 1b).

*Mul/modal UI.* Once the user will be trained and ready to use MI as browsing controls, the PAR-based menu configuration will be integrated with "MI shortcuts", that is, the possibility for the user to execute a MI to directly choose one of the available options. For example, if the user was trained in right hand MI, this task could be used to access communication mode (Speller from the menu in Fig. 1) and the confirmation phase could be skipped. The more MI the subject learns to use (and the model learns to recognize), the more shortcuts can be integrated in the PAR-based menu configuration.

*Simple answers.* Interaction with others is based on simple questions that require only a confirmation or denial answer from the patient. Therefore this design includes the possibility to trigger, via an external input (i.e. a push button), a special menu with only three choices: Yes, No and Don't want to answer. The selection follows the PAR-based menu paradigm.

We have to emphasise that the sketches proposed in Fig. 1 will then be shown and discussed with stakeholders such as doctors, caregivers and patients in order to review and redesign them in a co-design perspective [7]. Once implemented, the proposed UI will be tested in the wild with neurotypical users and then patients by proposing a set of gamified activities, as already successfully experienced in [12, 26], to make the experience more meaningful and enjoyable, and collect then feedback in a real context of use.

## 3.3 Neurofeedback Training

As anticipated above, Nti user training is a fundamental block for SMR-based BCIs and, given the subject-tailored nature of the application to be developed, its importance in this context gains even more room. It consists of a closed loop system, whose actors are i) the user, reproducing the required task, ii) the acquisition system, which streams real-time EEG data to iii) the classifier running on the PC, which in turn makes a prediction and gives it back to the user, through iv) an audio and/or visual apparatus [30]. The feedback reflects how close the user is to the ideal execution of the task, therefore allowing him/her to try different strategies to reach better or faster task recognition.

Being in this context Nti training sessions closely bounded to classifier adaptation, the interface used in this function will be also inspired by [11]. The main difference with the paradigm developed here stays in the number of tasks to be trained at once: Tireer et al. implemented a 4-classes training, while this work tends towards a more gradual path, as suggested in [2, 27], training a new task only after the user gets confident with the previous ones. Moreover, in this work the MI performance of the user must be tracked in order to assess the potential reliability of using it as control signal. According to [14], novel Riemannian-geometry based user performance metrics reflecting class separability and within-class consistency could be a valid index of user training progress, therefore these could be implemented and evaluated to define the switch from *PAR-based UI* to *Mul/modal UI*.

## 4 CONCLUSION

The vision presented in this short paper is essentially an attempt to merge the work done with PAR in [6] with the advances in adaptive MI classifiers represented here by [11] and [14], in a novel multimodal BCI application tailored on the specific user, following his/her progress in training those skills (MI execution in this preliminary prototype) which could restore, at least partially, his/her independence. The novelty of this work resides mainly in the use of PAR as additional control signal: this gives the user the possibility to interact with the system since the first moment with no need of training and through a quite natural act. Moreover, PAR control signal features the possibility to expand its communication potential in different ways. As done in [6], evaluating the duration of pupillary constriction may allow to define different commands. Alternatively, using a secondary display as near target would allow to think about new interface designs and increase the interaction speed. However, considering that CLIS patients cannot move the eyes, the two displays should be superimposed, so the near one should be semi-transparent. Moreover, PAR decoding algorithm should be tailored in order to be able to recognize single gaze shift and not complete PAR tasks as described in 3.1. Future works may head in this direction, with the objective of obtaining a smooth user experience involving the main stakeholders in the co-design and evaluation of both the UI and of the UX, matching CLIS patients needs and thus enabling an easier and effective interaction with the outer world.

ACKNOWLEDGMENTS

This research was funded by Fondazione CRT[4] in the context of the 2021 funding program, grant number 2021.0609.

---